\documentclass [11pt]{article}

\usepackage{amsmath}
\usepackage{amsfonts}
\usepackage{amssymb}
\usepackage{graphicx}
\usepackage{color}
\usepackage{ulem}

\usepackage{graphicx}
\usepackage[body={6.5in, 9in}, right=1in, top=1in]{geometry}
\usepackage{amssymb}
\usepackage{amsmath} 
\usepackage[numbers,sort&compress]{natbib}

\makeatletter
\def\section{\@startsection {section}{1}{\z@}{-3.5ex plus -1ex minus
 -.2ex}{2.3ex plus .2ex}{\large\bf}}
\def\subsection{\@startsection{subsection}{2}{\z@}{-3.25ex plus -1ex
minus -.2ex}{1.5ex plus .2ex}{\normalsize\bf}}
\makeatother
\makeatletter

\@addtoreset{equation}{section}

\makeatother

\def\be{\begin{equation}}
\def\ee{\end{equation}}

\newcommand\eea{\end{eqnarray}}
\newcommand\bea{\begin{eqnarray}}

\newcommand{\nn}{\nonumber}

\def\({\left(}
\def\){\right)}

\newcommand{\Comment}[1]{{}}
\definecolor{MyDarkBlue}{rgb}{0.15,0.15,0.45}
\usepackage[linktocpage=true]{hyperref}
\hypersetup{
colorlinks=true,
citecolor=MyDarkBlue,
linkcolor=MyDarkBlue,
urlcolor=MyDarkBlue,
}

\begin{document}
\def\thefootnote{\fnsymbol{footnote}}

\begin{center}
\Large{\textbf{Non-linear Representations of the Conformal Group  \\ [0.1cm]and Mapping of Galileons}} \\[0.5cm]
 
\large{Paolo Creminelli$^{\rm a}$, Marco Serone$^{\rm b,a,c}$, and Enrico Trincherini$^{\rm d,e}$}
\\[0.5cm]

\small{
\textit{$^{\rm a}$ Abdus Salam International Centre for Theoretical Physics\\ Strada Costiera 11, 34151, Trieste, Italy}}

\vspace{.2cm}

\small{
\textit{$^{\rm b}$ SISSA, via Bonomea 265, 34136, Trieste, Italy}}

\vspace{.2cm}

\small{
\textit{$^{\rm c}$ INFN - Sezione di Trieste, 34151 Trieste, Italy}}

\vspace{.2cm}

\small{
\textit{$^{\rm d}$ Scuola Normale Superiore, piazza dei Cavalieri 7, 56126, Pisa, Italy}}

\vspace{.2cm}

\small{
\textit{$^{\rm e}$ INFN - Sezione di Pisa, 56100 Pisa, Italy}}

\vspace{.2cm}

\end{center}

\vspace{.8cm}

\hrule \vspace{0.3cm}
\noindent \small{\textbf{Abstract}\\
There are two common non-linear realizations of the 4D conformal group: in the first, the dilaton is the conformal factor of the effective metric $\eta_{\mu\nu} \,e^{-2 \pi}$; in the second it describes the fluctuations of a brane in AdS$_5$. The two are related by a complicated field redefinition, found in \cite{Bellucci:2002ji} to all orders in derivatives.  We show that this field redefinition can be understood geometrically as a change of coordinates in AdS$_5$. In one gauge the brane is rigid at a fixed radial coordinate with a conformal factor on the AdS$_5$ boundary, while in the other one the brane bends in an unperturbed AdS$_5$. This geometrical picture illuminates some aspects of the mapping between the two representations. We show that the conformal Galileons in the two representations are mapped into each other in a quite non-trivial way: the DBI action, for example, is mapped into a complete linear combination of all the five Galileons in the other representation. We also verify the equivalence of the dilaton S-matrix in the two representations and point out that the aperture of the dilaton light-cone around non-trivial backgrounds is not the same in the two representations.  
} 
\vspace{0.3cm}
\noindent
\hrule
\def\thefootnote{\arabic{footnote}}
\setcounter{footnote}{0}

\section{Introduction}
The non-linear realization of symmetries is a cornerstone of modern quantum field theory. While the seminal papers \cite{Coleman:1969sm,Callan:1969sn} treated non-linearly realized internal symmetries, the extension to space-time symmetries was studied in \cite{Volkov:1973vd,Ivanov:1975zq}. The main qualitative difference between the two cases is that for space-time symmetries the number of Goldstones is less than the number of broken generators  (the so-called ``inverse Higgs phenomenon" \cite{Ivanov:1975zq}). In this paper we study the non-linear realization of the 4-dimensional conformal group, SO(4,2), where a single Goldstone appears, the dilaton.

Symmetries can be non-linearly realized on the fields of the theory in various ways, depending on how the coset space is parametrized. In the case of the conformal group two possibilities stand out. One is the representation constructed via the effective metric $g_{\mu\nu} = e^{-2 \pi} \eta_{\mu\nu}$. A covariant action for the effective metric $g$ non-linearly realizes SO(4,2), if the dilaton $\pi$ transforms in such a way to reabsorb the conformal factor induced by a Weyl transformation of the ordinary metric. We will refer in the following to this non-linear realization of the conformal group as the ``Weyl" representation.

A second representation emerges naturally in the context of the AdS/CFT correspondence where SO(4,2) appears geometrically as the isometry group of the 5-dimensional AdS space. An extended object (brane) at fixed radial position in AdS$_5$ breaks the conformal group to Poincar\'e. The scalar describing the brane position non-linearly realizes the SO(4,2) group, but in a way which differs from the Weyl representation. We will 
call this non-linear realization the ``DBI" representation, since the Dirac-Born-Infeld (DBI) action of the brane represents the simplest operator in this representation. The Weyl and DBI representations are related by an involved 
redefinition of the fields and the coordinates, remarkably found in \cite{Bellucci:2002ji} to all orders in a derivative expansion, through a generalization of the standard coset construction.

The aim of this paper is to study the physical properties of the mapping between the two representations. First of all, in Section \ref{sec:coord}, we will show that this mapping can be understood in a geometric way as a change of coordinates in AdS$_5$. Starting from the DBI representation, where the brane fluctuates in the unperturbed AdS$_5$ metric, one can go to a gauge where the brane is rigid at a fixed radial coordinate. This induces a conformal factor in the metric on the boundary of AdS$_5$ that precisely corresponds to the dilaton $\pi$ of the Weyl representation. In Section \ref{Sec:constr} we review the coset construction of the Weyl and DBI representations and their relation, following \cite{Bellucci:2002ji}.

A field redefinition, even if it involves a field dependent coordinate change, will lead to the same S-matrix scattering amplitude in Minkowski space:  indeed this is what happens in the case at hand, as we verify in few examples in Section \ref{sec:Smatrix}. The mapping of the two theories become more interesting when we are not interested in scattering elements, but in non-linear classical solutions and perturbations around them. 
This is the case for conformal operators which give equations of motion up to second order in derivatives: the conformal  Galileons \cite{Nicolis:2008in,deRham:2010eu}, considered in  Section \ref{sec:Galileons}.
Their interest lies in the possibility of studying in the regime of validity of the Effective Field Theory (EFT) non-linear solutions that lead to interesting modifications of gravity \cite{Nicolis:2008in} and novel cosmological evolutions (see for instance \cite{Nicolis:2009qm,Creminelli:2010ba,Hinterbichler:2012yn}). 
There are two sets of conformal Galileons, depending on which representation (Weyl or DBI) we are using. We show that the mapping of the two representations sends conformal Galileons in the Weyl representation (denoted Weyl Galileons for short in the following)  into conformal Galileons in the DBI representation  (denoted DBI Galileons from now on) and viceversa. This is quite easy to understand using our geometric view of the mapping as a change of coordinates in AdS$_5$, since a change of gauge cannot modify the property of Galileons of having second order equations of motion. The mapping is very non trivial since even the simplest operators in the Weyl representation, the kinetic dilaton and potential terms, are both mapped into a combination of all the five DBI conformal operators. Similarly the minimal DBI action is mapped into all the five Weyl Galileons. From this standpoint all the conformal Galileons appear much less ``exotic", as they can all be obtained from the simplest operators going in the other representation.

The mapping of non-trivial solutions raises some issues about super-luminality, defined in terms of the Minkowski light-cone,
as we discuss in Section \ref{sec:lightcones}. We will see that solutions whose perturbations are strictly subluminal, 
and therefore considered healthy, are mapped into solutions whose perturbations are on the verge of super-luminality, that would be considered pathological. (In particular the two Genesis scenarios of \cite{Creminelli:2010ba} and \cite{Hinterbichler:2012yn} are mapped into each other.) This raises the issue of how to interpret the constraint of absence of superluminality \cite{Adams:2006sv}. A complete answer lies beyond the scope of this paper and we hope we will come back to it in the near future. Conclusions and possible developments are discussed in Section \ref{sec:conclusions}.


\section{\label{sec:coord}The AdS Change of Coordinates}

The spontaneous breaking of the conformal group SO(4,2) to Poincar\'e is usually described in two different representations.\footnote{In this paper we focus on the 4D conformal group only, but 
our considerations can straightforwardly be extended to other space-time dimensions.}
The Weyl representation is the standard non-linear realization of dilatations and special conformal transformations in terms of the dilaton $\pi(x)$:
\bea
\label{diltransfD} \delta \pi_{D} & =& (1- x^\mu \partial_\mu \pi)c \,, \\
\label{diltransfK} \delta \pi_{K_\mu} &=& (-2x_\mu - x^2 \partial_\mu \pi +2 x_\mu x^\nu \partial_\nu \pi )b^\mu \,,
\eea
with $c$ and $b_\mu$ being the parameters of the infinitesimal transformations.
An action for $\pi$ which is invariant under the conformal group is conveniently written in terms of curvature invariants built out of the effective metric 
$g_{\mu\nu} = e^{-2 \pi} \eta_{\mu\nu}$. 

In the DBI representation it is useful to think of SO(4,2) as the group of isometries of AdS$_5$. In the presence of a probe 3-brane the subgroup ISO(3,1) is linearly realized, while the other isometries are broken. If we have a non-dynamical AdS$_5$ background
\be
\label{eq:AdS}
ds^2 = \frac{L^2}{z^2} (d x_\mu dx^\mu +dz^2)
\ee
and a brane in the position
\be
\bar z (x) = L \; e^{q(x)/L} \;,
\label{zbarL}
\ee
the leading order brane action is given by the usual Nambu-Goto action
\be
S_{NG} = - \frac{1}{L^4} \int d^4x \, e^{-4q/ L}\bigg( \sqrt{1+e^{2q/ L} (\partial q)^2} - 1\bigg)\,.
\label{SNG}
\ee
The branon field $q(x)$ can be seen in this case as the Goldstone boson of the broken transformations  \cite{Maldacena:1997re}, with $S_{NG}$
being the leading order terms in an expansion in invariants.
The non-linearly realized isometries of SO(4,2) act on $q(x)$ as:
\bea
\delta q_{\hat D} &=& \Big(1-\frac{1}{L} x^\mu \partial_\mu q\Big)c \,, \\
\delta q_{\hat K_\mu} &=& \Big(-2x_\mu - L  \, \partial_\mu q (e^{2q/L}-1)- \frac{1}{L} x^2 \partial_\mu q + \frac{2}{L} x_\mu x^\nu \partial_\nu q \Big)b^\mu \,.
\label{qeqtran}
\eea
Notice that the second term in (\ref{qeqtran}) does not appear in the Weyl representation (\ref{diltransfK}).
How are these two representations connected? The most intuitive answer can be given by thinking in terms of the rules of the AdS/CFT correspondence.
In the absence of any brane, an isometry of AdS can be seen as a conformal transformation on its boundary at $z=0$.
As said before, the presence of a brane in AdS breaks spontaneously some of its isometries, with the branon $q$ being the corresponding Goldstone field. 
We can now look for a change of coordinates $(x^\mu,z) \to (y^\mu,w)$ such that the brane in the new coordinates is at fixed $w$ and the boundary 4D metric
is conformally flat (this was analyzed at the linear level in \cite{Rattazzi:2000hs} and neglecting higher derivative terms in \cite{Elvang:2012st}).
In the new coordinates the asymptotic conformal factor of the 4D metric will play the role of the dilaton $\pi$. 
In other words, such a change of coordinates gives the relation between the two representations and trades the branon field $q$ for the dilaton $\pi$ in the Weyl representation. 

 Let us work it out explicitly: we want to perform the diffeomorphism $(x^\mu,z) \to (y^\mu,w)$ such that in the new coordinates the brane is at fixed $w$
\be
\bar w(y) = L
\ee
and with gauge conditions 
\be
\label{eq:gaugefix}
g_{\mu 5} =0\,, \qquad g_{55} = L^2/w^2 \,.
\ee
Consider a change of coordinates
\be
x^\mu = y^\mu + F^\mu(y,w)\,, \qquad z = w \,e^{G(y,w)} \,,
\label{CoordChange}
\ee
with
\be\label{CoordChange2}
F^\mu = -\frac{w^2}{2} e^{G(y,w)+\pi(y)}\eta^{\mu\nu} \partial_\nu\pi(y)\,, \qquad G = \pi(y) - \log\left(1 + w^2 \frac{e^{2\pi(y)}}{4} (\partial\pi(y))^2\right)\,.
\ee
The function $\pi(y)$ is arbitrary for the moment. It is straightforward to check that, independently of the choice of $\pi(y)$, this change of coordinates satisfies the gauge conditions \eqref{eq:gaugefix}. The first condition reads
\be
\frac{\partial F^\mu}{\partial w} + \frac{\partial F_\nu}{\partial w} \frac{\partial F^\nu}{\partial y^\mu} + w e^{2G} \left(1+ w \frac{\partial G}{\partial w}\right) \frac{\partial G}{\partial y^\mu} = 0 \;,
\ee
while the second becomes
\be
e^{-2 G} \left(\frac{\partial F^\mu}{\partial w}\right)^2 + 2 w \frac{\partial G}{\partial w} +  \left(w \frac{\partial G}{\partial w}\right)^2 = 0 \;.
\ee
The function $\pi(y)$ is fixed by the requirement that the brane is now at constant $w$: $\bar w(y) = L$. Using (\ref{zbarL}) and (\ref{CoordChange}), this condition reads
\be
L \, e^{q(x)/L} = L \,e^{\pi(y)} \left(1+\frac{L^2}{4} e^{2\pi(y)} (\partial\pi)^2\right)^{-1} \;.
\label{FieldMapping}
\ee
The metric in the new coordinates reads
\be
ds^2 = \frac{L^2}{w^2}\Big(g_{\mu\nu}(y,w) \,dy^\mu dy^\nu + dw^2\Big)\,.
\label{metricDual}
\ee
Close to the boundary $w=0$, the metric $g_{\mu\nu}$ can be expanded as
\be
g_{\mu\nu} = \eta_{\mu\nu} e^{-2\pi(y)} +{\cal O}(w^2) \;.
\ee
Therefore $\pi$ is the asymptotic conformal factor of the 4D metric in the new coordinates and transforms as the dilaton in (\ref{diltransfD}) and (\ref{diltransfK}).
It also corresponds to the radion, when we truncate the AdS space by a UV brane at $z=z_0$.  

This geometric picture makes evident the origin of the relation between the Weyl and DBI representations but somehow it does not explain how the change of coordinates (\ref{CoordChange}), (\ref{CoordChange2}) can be found.
In the next Section, following \cite{Bellucci:2002ji}, we will apply the technique of the coset construction to the case of SO(4,2) broken to Poincar\'e in the two different representations of the conformal algebra. The same coset manifold will be parametrized in terms of  space-time coordinates and Goldstone fields in two different ways and then by equating the two Cartan forms we will get explicitly the relation between the two set of coordinates.       

The reader interested in the application of this equivalence to the special case of Galileons can jump directly to Section \ref{sec:Galileons} where the mapping between DBI and Weyl Galileons is derived.

\section{Coset Construction}
\label{Sec:constr}

In this Section we review the coset construction of the two representations and their relation: we will follow closely \cite{Bellucci:2002ji} to which we refer for further details\footnote{In this paper we use the $(-,+,+,+)$ signature, while Bellucci, Ivanov and Krivonos \cite{Bellucci:2002ji} use $(+,-,-,-)$. Some additional change of notation:
\be
q = \frac{1}{\sqrt{2}} q_{\rm BIK} \,, \quad \pi = \Phi_{\rm BIK}\,, \quad L = \frac{1}{\sqrt{2} \;m_{\rm BIK}} \;.
\ee
}.
The Weyl representation is defined by the coset element
\be
g = e^{y^\mu P_\mu} e^{\pi D} e^{\Omega^\mu K_\mu} \;,
\ee
where $P_\mu$, $D$ and $K_\mu$ are the standard generators of the conformal group. The DBI representation is on the other hand defined by the coset
\be
g = e^{x^\mu P_\mu} e^{q \hat D} e^{\Lambda^\mu \hat K_\mu}\,,
\ee
where
\be
\hat K_\mu \equiv \frac{1}{\sqrt{2}L} K_\mu + \frac{L}{\sqrt{2}} P_\mu \,, \ \ \ \hat D \equiv  \frac{1}{\sqrt{2}L} D \;.
\ee
Going through the coset construction we get the Cartan form in the two representations. In the Weyl one we have
\be
\begin{split}
g^{-1} dg = & \, e^{-\pi} d y^\mu P_\mu + (d \pi - 2 e^{-\pi} \Omega_\mu d y^\mu) D - 4 e^{-\pi} \Omega ^\mu d y^\nu M_{\mu\nu} +\\
& \Big(d \Omega^\mu - \Omega^\mu d \pi + e^{-\pi} (2 \Omega_\nu d y^\nu \Omega^\mu - \Omega^2 dy^\mu)\Big) K_\mu \,,
\label{Coset1}
\end{split}
\ee
with $M_{\mu\nu}$ the Lorentz generators\footnote{Here and in the following, all indices are raised and lowered with $\eta_{\mu\nu}$.}. We can set to zero the expression multiplying the dilatation generator $D$ in (\ref{Coset1}) by imposing the so called  inverse Higgs constraint \cite{Ivanov:1975zq}. In this way we fix $\Omega_\mu$:
\be
\Omega_\mu(y) = \frac12 e^\pi \partial_\mu\pi(y) \;,
\ee
where the derivative is with respect to the coordinates $y^\mu$.
In the DBI representation we have
\be
\begin{split}
g^{-1} dg = & \bigg[e^{-q/L}\Big(dx^\mu - \frac{2\lambda^\mu \lambda_\nu dx^\nu}{1+\lambda^2}\Big)+\frac{2\lambda^\mu dq}{1+\lambda^2} \bigg] P_\mu +  \\
& \frac{1-\lambda^2}{1+\lambda^2}\bigg[ dq-2\frac{e^{-q/L}\lambda_\mu dx^\mu}{1-\lambda^2} \bigg] \sqrt{2} \hat D +  \\
&\frac{1}{1+\lambda^2} \bigg[d\lambda^\mu -\frac 1L \lambda^\mu dq -\frac{e^{-q/L}}L (\lambda^2 dx^\mu- 2\lambda^\mu \lambda_\nu dx^\nu) \bigg] \sqrt{2} \hat K_\mu +  \\
& \frac{2e^{-q/L}}{1+\lambda^2}\bigg[\frac 1L (\lambda^\nu dx^\mu - \lambda^\mu dx^\nu) + e^{q/L}(\lambda^\nu d\lambda^\mu - \lambda^\mu d\lambda^\nu)\bigg] M_{\mu\nu}\,,
\label{Coset2}
\end{split}
\ee
where
\be
\lambda_\mu = \Lambda_\mu \frac{\tan (\Lambda/\sqrt{2})}{\Lambda/\sqrt{2}}\,, \ \ \ \Lambda =\sqrt{\Lambda^\mu \Lambda_\mu}\,.
\ee
The inverse Higgs constraint gives now
\be
\lambda_\mu(x) = \frac{\partial_\mu q(x) \, e^{q(x)/L}}{1+\sqrt{1+e^{2q(x)/L} (\partial q(x))^2}}\,.
\ee
By equating the Cartan forms one finds the relation between the two representations
\be
y^\mu = x^\mu+ L e^{q(x)/L} \lambda^\mu(x)\,, \ \ \ \ \pi(y) = \frac{q(x)}L + \log(1+\lambda^2(x))\,, \ \ \  \Omega_\mu(y) = \frac 1L \lambda_\mu(x) \,.
\label{TransfEqs}
\ee
It is straightforward to check that the first and second relations in (\ref{TransfEqs})  coincide respectively with (\ref{CoordChange}) and (\ref{FieldMapping}) evaluated at $w=L$.

It is useful to have formulas which relate the coset constructions to more standard geometric tensors. In the Weyl representation, one defines the covariant derivative of the Goldstone \cite{Bellucci:2002ji}
\be
{\cal D}_\nu \Omega_\mu = \frac{e^{2\pi}}{2}\Big(\partial_\mu\partial_\nu\pi + \partial_\mu\pi\partial_\nu\pi-\frac12 (\partial\pi)^2 \eta_{\mu\nu}\Big)\,.
\label{DOmega}
\ee
One can thus write the Ricci tensor of the effective metric  $g_{\mu\nu} = \eta_{\mu\nu} e^{-2\pi}$ as
\be
e^{2\pi} R_{\mu\nu}(g) = 4 {\cal D}_\mu \Omega_\nu + 2 \eta_{\mu\nu} D^\alpha \Omega_\alpha \,.
\label{RmunuOmega}
\ee
In the DBI representation, the covariant derivative reads
\be
{\cal D}_\mu\lambda^\nu =  \frac{1}{1+\lambda^2} \bigg(e^{q/L}\Big(\partial_\mu \lambda^\nu-2\frac{\lambda_\mu \lambda^\rho\partial_\rho \lambda^\nu}{1+\lambda^2}\Big) - \frac 1L \lambda^2 \delta_\mu^\nu\bigg)\,.
\ee
In the $q$ coordinates, the AdS metric (\ref{eq:AdS}) reads
\be
ds^2 = e^{-2q/L} dx^\mu dx_\mu + dq^2 \,.
\label{AdSqmetric}
\ee
From (\ref{Coset2}) one gets the brane induced vierbein and metric, and their inverses:
\bea
E_\nu^\alpha  & = & e^{-q/L}  \Big(\delta_\nu^\alpha+2\frac{\lambda_\nu\lambda^\alpha}{1-\lambda^2}\Big)\,, \ \ \ \ 
G_{\mu\nu} =  E_\mu^\alpha E_\nu^\beta \eta_{\alpha\beta} =  e^{-2 q/L}  \eta_{\mu\nu}+\partial_\mu q \partial_\nu q\,, \nn \\ 
 (E^{-1})_\mu^\alpha & = &  e^{q/L} \Big(\delta_\mu^\alpha-2\frac{\lambda_\mu\lambda^\alpha}{1+\lambda^2}\Big)\,, \ \ \  \ \  \
 G^{\mu\nu}  =  e^{2q/L}\eta^{\mu\nu} - \frac{e^{4q/L}\partial^\mu q \partial^\nu q}{1+e^{2q/L}(\partial q)^2}\,.
 \label{GmunuInd}
\eea
The extrinsic curvature in curved space reads
\be
K_{\mu\nu} = \frac{\partial X^A}{\partial x^\mu}  \frac{\partial X^B}{\partial x^\nu} \nabla_A n_B\,,
\label{ExtDef}
\ee
where $n_A$ is a vector orthonormal to the surface, namely
\be
 \frac{\partial X^A}{\partial x^\mu} n^B \hat G_{AB} = 0\,, \ \ \ \  n^A n^B \hat G_{AB} = 1\,,
 \label{Ortho}
\ee
with $\hat G$ the 5D AdS metric (\ref{AdSqmetric}).
In (\ref{ExtDef}) and (\ref{Ortho}), $X^A$ is the brane embedding vector. In the static gauge we take it to be 
\be
X^A=(x^\mu, q(x))\,.
\ee
Explicitly, we find
\be
\begin{split}
K_{\mu\nu} & =  -\frac{1}{\sqrt{1+e^{2q/L}(\partial q)^2}} \Big( \partial_\mu\partial_\nu q + \frac 1L \partial_\mu q \partial_\nu q +\frac 1L G_{\mu\nu}\Big) \,, \\
E_\mu^\sigma E_\nu^\rho {\cal D}_\sigma \lambda_\rho  & = - \frac 12 \Big( K_{\mu\nu} +\frac{1}{L} G_{\mu \nu}\Big)\,.  \label{Klambda}
\end{split}
\ee

Finally, we report useful formulas relating the Weyl and DBI  representations: 
\be
\begin{split}
\frac{\partial y^\nu}{\partial x^\mu}  = & e^{q/L} (1+\lambda^2)  E^\rho_\mu   (\delta^\nu_\rho + L {\cal D}_\rho \lambda^\nu) \equiv e^{q/L}  (1+\lambda^2)  E^\rho_\mu T_\rho^\nu\,,  \\
T^\nu_\rho = & \frac12\delta^\nu_\rho -\frac{L}2 K_{\alpha \beta} (E^{-1})^\alpha_\rho (E^{-1})^{\beta\nu} \label{TKmunu}\,,
\end{split}
\ee
and the important relations between the covariant derivatives:
\bea
{\cal D}_\nu \Omega_\mu & = &  \frac1L (T^{-1})^\omega_\nu {\cal D}_\omega \lambda_\mu \,,
\label{Omega2lambda} \\
{\cal D}_\nu \lambda_\mu & = & L T^\omega_\nu {\cal D}_\omega \Omega_\mu\,.
\label{lambda2Omega}
\eea
Thanks to (\ref{Omega2lambda}) and (\ref{lambda2Omega}), and (\ref{RmunuOmega}), (\ref{Klambda}), relating covariant derivatives to geometric tensors, 
we can directly map geometric invariants from one representation to the other. We will see this map in some more detail for the relevant case of the Galileons in the next Section.

\section{\label{sec:Galileons}Galileon Mapping}

So far our discussion has been general and valid for any possible conformal action. We now focus on a particular set of five operators in each representations: the conformal Galileons.  We will show that, in going from one representation of the conformal group to the other, the five Galileons are mapped into themselves: each Weyl Galileon is mapped into a linear combination of the DBI Galileons and viceversa. 

Let us start by introducing the two sets of operators. The Weyl Galileons were introduced in \cite{Nicolis:2008in} as a natural extension of the Galilean symmetry to the conformal group (in the Weyl representation). The Weyl Galileons are particular linear combinations of the conformal operators with $2 n$ derivatives in which terms of the form $(\partial\partial\pi)^n$ combine to give total derivatives 
and have second order equations of motion.

There are only five Weyl Galileons in 4D:
\bea  
{\cal L}_{\pi1}&=&-e^{-4\pi} \ , \nn\\
{\cal L}_{\pi2}&=&- L^2 e^{-2\pi}(\partial\pi)^2 \ ,\nn \\
{\cal L}_{\pi3}&=& L^4(\partial {\pi})^2\(- [\Pi]+\frac{1}{2} (\partial {\pi})^2 \)\ , \label{DBIgalderivexpan}\\
{\cal L}_{\pi4}&=& L^6 e^{2 \pi}(\partial  \pi)^2
\(-[ \Pi]^2+[ \Pi^2]-{1\over 2}(\partial  \pi)^2[\Pi]-{1\over 2}(\partial  \pi)^4\) \ , \nn \\
{\cal L}_{\pi5}&=& L^8 e^{4 \pi} (\partial  \pi)^2
\Big[- [ \Pi]^3+3[ \Pi][ \Pi^2]-2[ \Pi^3]-3(\partial  \pi)^2([ \Pi]^2-[ \Pi^2]) \nn
-5 (\partial  \pi)^4[ \Pi]-\frac{11}{4}(\partial  \pi)^6 \Big] \ .
\eea
Some explanation of the notation is in order. $\Pi$ is the matrix of second derivatives $\Pi_{\mu\nu}\equiv\partial_{\mu}\partial_\nu\pi$.  For traces of the powers of $\Pi$ we write $[\Pi^n]\equiv {\rm Tr}(\Pi^n)$, e.g. $[\Pi]=\partial_\mu\partial^\mu\pi$, $[\Pi^2]=\partial_\mu\partial_\nu\pi\partial^\mu\partial^\nu\pi$.  We define the contractions of the powers of $\Pi$ with $\partial\pi$ using the notation $[\pi^n]\equiv \partial\pi\cdot\Pi^{n-2}\cdot\partial\pi$, e.g. $[\pi^2]=\partial_\mu\pi\partial^\mu\pi$, $[\pi^3]=\partial_\mu\pi\partial^\mu\partial^\nu\pi\partial_\nu\pi$.\footnote{It is useful to note the following total derivative
\be
\partial_\mu \left[e^{4\pi} (\partial\pi)^6 \partial^\mu\pi\right] = e^{4\pi} (\partial\pi)^2 \left[ \Box\pi (\partial\pi)^4 + 6 [\pi^3] (\partial\pi)^2 +4 (\partial\pi)^6 \right] \;.
\ee
If we add $5 L^8/7$ of this to ${\cal L}_{\pi5}$ we get the same form as given in \cite{deRham:2010eu,Creminelli:2012my} up to the overall normalization.
} Powers of $L$ have been introduced in (\ref{DBIgalderivexpan}) to make the operators dimensionless.
With the exception of ${\cal L}_{\pi3}$, the Weyl Galileons can also be written in terms of the metric
\be
g_{\mu\nu} = \eta_{\mu\nu} e^{-2\pi}  
\label{gmunupi}
\ee
and its curvature:\footnote{These expressions will coincide with \eqref{DBIgalderivexpan} up to total derivatives.}
\be
\begin{split}
{\cal L}_{\pi1} = &  -\sqrt{-g}\,, \\
{\cal L}_{\pi2}  =  & - L^2\frac{\sqrt{-g}}6 R\,,  \\
{\cal L}_{\pi4}=  & - L^6\frac{\sqrt{-g}}{4} \left(-\frac{7}{36}R^3 + R (R_{\mu\nu})^2- (R_{\mu\nu})^3\right)\,,  \\
{\cal L}_{\pi5}=  & \, L^8\frac{\sqrt{-g}}2 \left(\frac{93}{2 \cdot 6^4} R^4 - \frac{39}{4 \cdot 6^2} R^2  (R_{\mu\nu})^2 + \frac5{12} R  (R_{\mu\nu})^3 +\frac3{16} (R_{\mu\nu}^2)^2 -\frac38 (R_{\mu\nu})^4  \right)\,.
\label{DilInv}
\end{split}
\ee
For ${\cal L}_{\pi3}$ an analogous expression only exists in $d \neq 4$ and one can only write ${\cal L}_{\pi3}$ as a $d \to 4$ limit \cite{Nicolis:2008in,Komargodski:2011vj}.

The DBI Galileons were introduced in \cite{deRham:2010eu}: they are all the operators in the DBI representation that preserve second order equations of motion. There are again five of them \cite{Goon:2011qf}:\footnote{The expressions of ${\cal L}_{q4}$ and ${\cal L}_{q5}$ in \cite{Goon:2011qf} contain typos that we corrected. We thank G.~Trevisan for help with this.}
\be
\begin{split}
{\cal L}_{q1}= &  -e^{-4q/ L} \,,\\
{\cal L}_{q2}= & -e^{-4q/ L}\sqrt{1+e^{2q/ L} (\partial q)^2} \,, \\
{\cal L}_{q3}= & \, L \, \gamma^2[q^3]- L\,e^{-2q/ L}[Q]+e^{-4q/ L}(\gamma^2-5) \,, \\
{\cal L}_{q4}= & 
L^2 \gamma([Q]^2-[Q^2])+2 L^2 \gamma^3 e^{2q/L}([q^4]-[Q][q^3]) \\
&-6 e^{-4q/L}{1\over \gamma}\(2-3\gamma^2+\gamma^4\)-8 L\gamma^3 [q^3]
+ 2 L e^{-2q/L}\gamma\(4-\gamma^2\)[Q]
 \,,  \\
{\cal L}_{q5}= & 
2 L^3 \gamma^2 e^{2q/ L}\([Q]^3-3[Q][Q^2]+2[Q^3]\)  \\ 
\hspace{-5pt}&\hspace{-5pt}
+6 L^3 \gamma^4e^{4q/ L}\left[2([Q][q^4]-[q^5])-([Q]^2-[Q^2])[q^3]\right]\\
\hspace{-5pt}&\hspace{-5pt} -36 L^2 e^{2q/ L}\gamma^4([Q][q^3]-[q^4]) + 6 L^2 {\gamma^2}(3-\gamma^2)([Q]^2-[Q^2]) \\
\hspace{-5pt}&\hspace{-5pt}+3 L { \gamma^2}(3-20\gamma^2)[q^3]
-3 L e^{-2q/ L}(3-20\gamma^2+8\gamma^4)[Q] \hspace{-100pt} \\
\hspace{-5pt}&\hspace{-5pt}-3 e^{-4q/ L}(15-31\gamma^2+12\gamma^4)
\ .\label{conformalDBIGalileonterms}
\end{split}
\ee
We use here the same notation as for the Weyl Galileons and
\be
\gamma \equiv {1\over \sqrt{1+e^{2q/ L} (\partial q)^2}} \ .
\ee
Notice that the NG action (\ref{SNG}) is given by a combination of the first two DBI Galileon terms in (\ref{conformalDBIGalileonterms}), ${\cal L}_{NG}= L^{-4} (-{\cal L}_{q1}+{\cal L}_{q2})$. 
Also in this case it is convenient to think about these operators in geometric terms \cite{deRham:2010eu, Goon:2011qf}. One writes operators on a probe brane in AdS$_5$, preserving second order equations of motion. They can be written in terms of the metric $G_{\mu\nu}$ induced on the brane (\ref{GmunuInd}),
\be
G_{\mu\nu} = e^{-2q/L} \eta_{\mu\nu}+\partial_\mu q\partial_\nu q\,.
\label{IndMetric}
\ee
This is their explicit form:
\be
\begin{split}
{\cal L}_{q1}=& -e^{-4q/ L} \,, \\
{\cal L}_{q2}=& -\sqrt{-G} \,, \\
{\cal L}_{q3}=& \, L \sqrt{- G} \; K \,, \\
{\cal L}_{q4}=&
- L^2 \sqrt{- G} \;R \, = - L^2 \sqrt{- G}  \left(\frac{12}{L^2} - [K]^2 + [K^2] \right)\,, \\
{\cal L}_{q5}=& \,
\frac{3 L^3}2 \sqrt{- G} \;K_{GB} = \frac{3 L^3} 2 \sqrt{-G} \left(6 K L^{-2} -\frac83 [K^3] + 4 [K] [K^2] -\frac43 [K]^3 \right)
\ .\label{CurvInvqi} 
\end{split}
\ee
Here $K_{\mu\nu}$ is the extrinsic curvature of the brane (\ref{Klambda}). The operator $K_{GB}$ is the boundary term associated to the Gauss-Bonnet term in the bulk \cite{deRham:2010eu}. In the last two equations we have written the operators in terms of the extrinsic curvature, using the Gauss-Codazzi relation.

At first, there is no obvious reason why the two sets of Galileons should be mapped into each other by the change of representation. The fact that they both give second order equations of motion does not help, since this property is not preserved under a general field redefinition. The AdS picture we developed in Section \ref{sec:coord}, on the other hand, clearly shows the link between the two sets of Galileons. The two representations are related by a change of coordinates: the property of having second order equation of motion cannot depend on the choice of coordinates. Given that the two sets of Galileons are the only operators with this property, we argue that the two sets must be mapped into each other.

It is however useful to verify the mapping in detail.  This is how the Weyl Galileons are written in terms of the DBI ones and viceversa:
\def\arraystretch{1.2}
\be
\label{Dil2AdS}
\left( \begin{array}{c}
{\cal L}_{\pi 1} \\ {\cal L}_{\pi 2} \\ {\cal L}_{\pi 3} \\ {\cal L}_{\pi 4} \\ {\cal L}_{\pi 5} 
\end{array} \right) =
\left(  \begin{array}{ccccc}
0 & \frac12 & \frac{7}{64} & - \frac1{24} & -\frac{1}{192} \\
0 & 0 & -\frac{1}{16} & -\frac{1}{12} & -\frac{1}{48}\\
4 & 0 & -\frac{11}{8} & 0 & - \frac18 \\
0 & 0 & -\frac32 & 2 & -\frac12 \\ 0 & -96 & 21 & 8 & -1
\end{array} \right)
\left( \begin{array}{c}
{\cal L}_{q 1} \\ {\cal L}_{q 2} \\ {\cal L}_{q 3} \\ {\cal L}_{q 4} \\ {\cal L}_{q 5} 
\end{array} \right) \,,
\ee

\be
 \label{AdS2Dil}
\left( \begin{array}{c}
{\cal L}_{q 1} \\ {\cal L}_{q 2} \\ {\cal L}_{q 3} \\ {\cal L}_{q 4} \\ {\cal L}_{q 5} 
\end{array} \right) =
\left(  \begin{array}{ccccc}
1 & -1 & \frac{1}{4} & - \frac1{24} & \frac{1}{192} \\
1 & -\frac12 & 0 & \frac{1}{48} & -\frac{1}{192}\\
4 & -1 & 0 & -\frac{1}{24} & \frac1{48} \\
0 & -6 & 0 & \frac14 & 0  \\ -12 & -21 & 0 & -\frac78 & -\frac1{16}
\end{array} \right)
\left( \begin{array}{c}
{\cal L}_{\pi 1} \\ {\cal L}_{\pi 2} \\ {\cal L}_{\pi 3} \\ {\cal L}_{\pi 4} \\ {\cal L}_{\pi 5} 
\end{array} \right)\,.
\ee

To get these relations we started from \eqref{DilInv}, the expression of the ${\cal L}_{\pi i}$'s in terms of curvature invariants. Using (\ref{RmunuOmega}), (\ref{Omega2lambda}), (\ref{Klambda}) and (\ref{TKmunu}) one is able to map these operators in the DBI representation and write them directly in terms of the extrinsic curvature $K_{\mu\nu}$. From this it is easy to get to the ${\cal L}_{qi}$'s, using their expression \eqref{CurvInvqi} in terms of the extrinsic curvature. Notice that in this way we always produce a determinant of the induced metric on the brane, so that we never generate the term ${\cal L}_{q1}$. This procedure works except for ${\cal L}_{\pi3}$, since its geometric expression would require a complicated $d \to 4$ limit. Instead, we prefer to fix this row of the matrix looking at the inverse transformation. 
We start from the definition of the ${\cal L}_{q i}$'s in terms of the brane geometry, \eqref{CurvInvqi}, and express them in terms of the  ${\cal L}_{\pi i}$ using  (\ref{Klambda}), (\ref{lambda2Omega}) and (\ref{RmunuOmega}), where in (\ref{lambda2Omega}) the matrix $T^\omega_\nu$ should be
seen as the inverse of
\be
(T^{-1})^\omega_\nu = \delta^\omega_\nu - L^2 {\cal D}_\nu\Omega^\omega \,.
\ee

This procedure does not work for ${\cal L}_{q1}$, but in this case we can directly use eqs.~(\ref{TransfEqs}) and express the result in terms of the Weyl Galileons (\ref{DBIgalderivexpan}). In this way we have derived the whole inverse map (\ref{AdS2Dil}). By computing its inverse, we have fixed the last unknown row of the map (\ref{Dil2AdS}) and checked that the remaining entries
of the matrices coincide. As a further check, we have also computed explicitly the transformation of ${\cal L}_{q2}$ and written it in terms of the Weyl Galileons (\ref{DBIgalderivexpan}): the result agrees with the expression obtained using the curvature invariants.  

It is important to point out a subtlety in the procedure above. The DBI Galileons correspond to the Lovelock invariants on the brane and the boundary terms associated to the Lovelock terms in the 5D bulk \cite{deRham:2010eu}. These are the only terms that guarantee second order equations of motion. If one continues the list of \eqref{CurvInvqi}, the following term, ${\cal L}_{q6}$, would be the Gauss-Bonnet term on the brane. This, however, is a total derivative. Written in terms of the extrinsic curvature it reads
\be
\begin{split}
\label{Lq6}
{\cal L}_{q6} =
L^4 \sqrt{- G} \;(R^2 -4 R_{\mu\nu}^2 + R_{\mu\nu\rho\sigma}^2) & = L^4 \sqrt{-G} \left(24 L^{-4} - 4 [K]^2 L^{-2} + 4 [K^2] L^{-2} \right. \\ & \left. -6 [K^2][K]^2 +3 [K^2]^2 + [K]^4 -6 [K^4] + 8 [K] [K^3]\right) \;.\\
\end{split}
\ee
If we start from the Weyl Galileons and go through the above procedure we will generate terms of the form $K^4$. These, by themselves, {\it do not} form a total derivative, but only when combined with lower order terms to give rise to \eqref{Lq6}. This has to be taken into account since it contributes to the coefficients of the other ${\cal L}_{qi}$'s. This subtlety exists only at order $K^4$, since Lovelock invariants of higher order are not total derivatives, but vanish identically. 
Notice that the same thing does not occur in the Weyl representation. There is no combination ${\cal L}_{\pi 6}$ of the schematic form $R_{\mu\nu}^5$ which is a total derivative. Indeed all terms of the form $(\partial^2\pi)^5$ must combine to form a total derivative in order to keep second order equations of motion. But, as explained in \cite{Nicolis:2008in}, there are no total derivatives of the form $(\partial^2\pi)^n$ with $n > 4$. Thus the terms $(\partial^2\pi)^5$ must cancel one by one and this implies that the whole linear combination of $R_{\mu\nu}^5$ terms vanishes. This also explains the fact that in going from the DBI to the Weyl representation all terms which are generated beyond the five ${\cal L}_{\pi i}$'s vanish identically, since it is not possible to write any total derivative in terms of the curvature tensor. The same thing  happens in the opposite direction, with the only exception of the ${\cal L}_{q6}$ we just discussed.  

Notice that in the map (\ref{Dil2AdS}) the term ${\cal L}_{q1}$  only contributes to ${\cal L}_{\pi3}$.
Viceversa, in the inverse map (\ref{AdS2Dil}), the term ${\cal L}_{\pi3}$ only contributes to ${\cal L}_{q1}$.
This is a manifestation of the fact that neither ${\cal L}_{\pi3}$ nor ${\cal L}_{q1}$ can be written in terms of curvature invariants.
The first has been shown to come from the Wess-Zumino term associated with the Weyl anomaly  \cite{Komargodski:2011vj}, the latter
come from Wess-Zumino couplings of D-branes in UV string realizations. See \cite{Goon:2012dy} for an interpretation of ${\cal L}_{q1}$ and ${\cal L}_{\pi 3}$ as Wess-Zumino terms associated with the coset construction reviewed in Section 3.

\section{\label{sec:Smatrix}Equivalence of the S-matrix}

The Weyl and DBI non-linear representations of the conformal group are related by (\ref{CoordChange}) and (\ref{FieldMapping}). 
The mapping (\ref{FieldMapping}), properly expanded in derivatives,  can be seen as
a particular (though highly non-trivial) implicit field redefinition which does not affect the space-time coordinates. 
Since the $S$-matrix is known to be invariant under such field redefinitions, on-shell scattering amplitudes should be the same in both representations.
We explicitly show this equivalence for the particular case of $2\rightarrow 2$ dilaton scattering around Minkowski.

In the DBI representation we start from the NG action (\ref{SNG}). Expanding in derivatives up to $(\partial q)^4$ terms, we get
\be
{\cal L}_{NG} = - \frac{1}{2} (\partial q)^2 +\frac 1{8 f^4} (\partial q)^4\,, 
\label{LNGExp}
\ee
where we have canonically normalized $q$ and have defined the dilaton decay constant
\be
f^2\equiv \frac{1}{L^2}\,.
\ee
From (\ref{LNGExp}) a straightforward computation gives, at tree-level, 
\be
{\cal A}_{DBI}(2\rightarrow 2) = \frac{s^2+t^2+u^2}{4f^4} \,,
\label{2to2}
\ee
where $s$, $t$, and $u$ are the usual Mandelstam variables. Notice that (\ref{2to2}) is tree-level exact, since higher order terms from the expansion of the square root in the NG action necessarily appear with more than four  dilaton fields.\footnote{This is of course an artifact of our choice of action. Starting from an effective action involving higher order
invariants will in general give rise to higher order corrections in (\ref{2to2}).}  The map (\ref{AdS2Dil}) gives
\be
{\cal L}_{NG} = \frac{1}{L^4} (-{\cal L}_{q1}+{\cal L}_{q2}) =\frac{1}{L^4} \Big(\frac 12  {\cal L}_{\pi2} -\frac 14 {\cal L}_{\pi3} +\frac{1}{16} {\cal L}_{\pi4} -\frac{1}{96} {\cal L}_{\pi5}\Big)\,.
\label{DBImapped}
\ee
By performing the field redefinition\footnote{The reader may think we are cheating since we show the equivalence of our complicated field redefinition \eqref{TransfEqs} using here another (much simpler) field redefiniton. On the other hand the equivalence of the S-matrix for these simple field redefinitions is well understood. Moreover, one can verify that the action \eqref{DBImapped} gives the amplitude \eqref{2to2} directly, without field redefinitions.}
\be
\pi \rightarrow \pi + \frac 12 \pi^2 -\frac 14 L^2 (\partial \pi)^2 
\label{fieldredef}
 \ee
all terms cubic in $\pi$ can be removed from the action (\ref{DBImapped}). Modulo $\Box \pi$ terms that vanish on-shell, and keeping terms involving no more than 4 dilatons, one gets back the action (\ref{LNGExp})
in terms of a canonically normalized dilaton field $\pi$. It then trivially follows that 
\be
{\cal A}_{DBI}(2\rightarrow 2) ={\cal A}_{Weyl}(2\rightarrow 2)\,.
\label{2to2Ex}
\ee
Other simple checks of the mapping  (\ref{FieldMapping}) can be performed. For instance, in the Weyl representation, the action ${\cal L}_{\pi 2}$ describes a free dilaton (this is easily seen by defining  $\Omega = 1-\exp(-\pi)$) and should map to a free theory as well in the DBI representation. Indeed, one can check that the $2\rightarrow 2$ amplitude vanishes.

In the Weyl representation, the $2\rightarrow 2$ dilaton scattering at low energies is governed by the ${\cal L}_{\pi3}$ term. Positivity of the total cross-section implies that the coefficient multiplying 
this term has to be negative and, by means of the map (\ref{Dil2AdS}), this implies that 
\be
c_{q1} < 0\,.
\ee
This bound is in particular respected in the NG action, where a definite positive kinetic term for $q$ and absence of a vacuum energy requires $c_{q1} = - c_{q2} = -1$.

Let us check the equivalence in the presence of external sources by considering the addition of a massless scalar field $\phi$. 
In the Weyl representation, the action reads
\be
S_{\pi+\phi} =  -\frac{1}{L^2}\int d^4y \,  e^{-2\pi} (\partial\phi)^2 \,.
\label{Lag_piphi}
\ee
At tree-level, the scattering $\phi \phi\rightarrow \phi \phi$ can only be induced by the exchange of a single $\pi$, coming from the expansion
of the exponential factor. By Bose symmetry, the amplitude is proportional to $s+t+u=0$ and is trivial. 

In the DBI representation the simple-looking Lagrangian turns into a complicated form
\be
S_{\pi+\phi}\rightarrow S_{q+\phi} = -\frac{1}{L^2} \int d^4x \, {\rm det}\,T(x) \, \frac{(1+\lambda(x)^2)^3}{1-\lambda(x)^2} e^{-2q(x)/L} (\partial_y \phi(y))^2\,,
\label{Lag_qphi}
\ee
The tree-level scattering $\phi \phi\rightarrow \phi \phi$ can still be induced by the exchange of a single dilaton. Expanding (\ref{Lag_qphi}) at leading order in $q$, we get
\be
S_{q+\phi}  = -\frac{1}{L^2} \int d^4x \Big( e^{-2q/L} (\partial \phi)^2 +\frac L2 \Box q   (\partial \phi)^2 + L \partial_\nu \phi \partial_\mu q \partial_\mu \partial_\nu \phi \Big) =  -\frac{1}{L^2} \int d^4x \, e^{-2q/L} (\partial \phi)^2 \,,
\label{Lag_qphi2}
\ee
since the last two terms combine in a total derivative. At this order, the action (\ref{Lag_qphi}) coincides with (\ref{Lag_piphi}) and results in the same trivial amplitude.
This simple exercise shows that the equivalence between the DBI and the Weyl representations hold in  presence of additional fields.

\section{\label{sec:lightcones}Lightcones}

The equivalence of the Weyl and DBI representations provided by our mapping presents also puzzling aspects, since it has been pointed out that the theories based on the former can lead to superluminal
propagation of fluctuations in certain backgrounds \cite{Nicolis:2009qm,Creminelli:2010ba}, while no superluminal propagation is possible in theories based on the NG action (\ref{SNG}).

Let us consider the propagation of fluctuations around Minkowski space, in the presence of a background configuration $q_0(x)$ of the form
\be
L\, \partial_\mu e^{q_0(x)/L} = C_\mu\,,
\label{Cback}
\ee
where $C_\mu$ is a constant vector. We assume that (\ref{Cback}) is a classical solution of the NG action (\ref{SNG}) with the addition of suitable sources.
For simplicity, we consider $C^2 \ll 1$.
Let us analyze the fluctuations of the canonically normalized field $\chi = e^{-q/L}/L$. Up to quadratic order in both the fluctuations and the background $C_\mu$, the NG action  (\ref{SNG}) reads
\be
S_{NG} = \int\!d^4x \Big(-\frac 12 (\partial \chi)^2 + \frac{C^2}4 (\partial\chi)^2 +\frac 12 (\partial\chi\cdot C)^2\Big) \,.
\label{NGchiExp}
\ee
The equation of motion of the fluctuations $\chi(x)$ coming from (\ref{NGchiExp}), modulo an overall constant rescaling,  is
\be
\Big(\eta^{\mu\nu} - C^\mu C^\nu \Big) \partial_\mu \partial_\nu \chi = 0\,.
\label{DBI_lightcone}
\ee
The second term in (\ref{DBI_lightcone}) implies that free plane waves of the DBI dilaton propagate strictly  sub-luminally around the background (\ref{Cback}), 
with respect to the Minkowski light-cone defined by $\eta_{\mu\nu}$.

The background (\ref{Cback}) has the nice property of being essentially invariant under the mapping (\ref{TransfEqs}). One has
\be
L\, \partial_\mu e^{\pi_0(y)} = \frac{2C_\mu}{1+\sqrt{1+C^2}} = C_\mu \Big(1+{\cal O}(C^2) \Big)\,.
\ee
Within the same approximations as above, the equation of motion for the fluctuations in the canonical field $\phi = e^{-\pi}/L$  reads
\be
\eta^{\mu\nu} \partial_\mu\partial_\nu \phi(y) = 0\,,
\label{dilaton_lightcone}
\ee
namely free plane waves of $\phi$ propagate at the speed of light, with respect to the Minkowski light-cone.
We see that the change of representation maps the Minkowski light-cone of the Weyl representation to the light-cone of the induced metric on the brane (see \eqref{DBI_lightcone}) in the DBI case which is, 
modulo an overall factor,
\be
G^{\mu\nu} = \eta^{\mu\nu} - C^\mu C^\nu + \ldots \,\,.
\ee
A relevant special case of (\ref{Cback}) is provided by the Genesis scenarios, based on Galileon operators (either in the Weyl  \cite{Creminelli:2010ba} or DBI representations \cite{Hinterbichler:2012fr,Hinterbichler:2012yn}) with an SO(4,1) invariant background, in which
\be
\label{eq:SO41back}
e^{\pi} \propto t \,,\qquad e^{q/L} \propto t \,.
\ee
The two solutions \eqref{eq:SO41back} are related by
\be
e^\pi = \alpha_\pi \cdot y^0 \quad \to \quad e^{q/L} = \alpha_q \cdot x^0   
\ee
with
\be
 \alpha_q = \alpha_\pi \left(1+\frac{L^2 \alpha_\pi^2}4\right)^{-1}  \quad {\rm and} \quad x^0 = y^0 \cdot \frac{1+\frac{L^2 \alpha_\pi^2}{4 }}{1-\frac{L^2 \alpha_\pi^2}{4 }} \;.
\ee
The time $x^0$ of the AdS parametrization is dilated compared to the dilaton one, $y^0$, so that the speed of propagation will be subluminal in the AdS case. Indeed in \cite{Hinterbichler:2012fr,Hinterbichler:2012yn} it is found that the symmetries in the AdS parametrization force a subluminal propagation by a factor $1/\gamma$ (the relativistic factor of the brane motion in AdS)
\be
\gamma^2 = \frac1{1-\frac{L^2}{\alpha_q^2}} \;.
\ee
This matches exactly with what we get starting from a luminal propagation in the $\pi$ variables and taking into account the time dilation above.
As argued in \cite{Nicolis:2009qm,Creminelli:2010ba} the luminal propagation of perturbations in the Weyl case is problematic. A small deformation of the solution is enough to allow superluminal propagation. This superluminality is measurable within the EFT, unless its regime of validity is limited to a scale lower than $1/t$, but in this case the solution itself cannot be trusted. In the DBI variables the light-cone is closed with respect to the Minkowski metric and a small deformation cannot induce any superluminality. On the other hand, the light-cone is null with respect to the induced brane metric.

 It is known that in presence of dynamical gravity the criterion of luminality with respect to the Minkowski light-cone cannot be used straightforwardly \cite{Adams:2006sv}, since the natural metric to use is $g_{\mu\nu}$. In our case gravity is decoupled but the examples above clearly show that a field-dependent change of coordinates can affect the light-cone in non-trivial backgrounds. 
 
Different couplings of the Galileons with dynamical gravity are possible in the two representations \cite{Nicolis:2009qm, Deffayet:2009wt,deRham:2010eu,Creminelli:2012my} and the issue is of particular importance when discussing the violation of the Null Energy Condition. It would be interesting to study how the different couplings transform under the mapping.

\section{\label{sec:conclusions}Concluding remarks}

We have analyzed in this paper various aspects of the field redefinition found in \cite{Bellucci:2002ji}, that relates two different non-linear realizations of the 4D conformal group.
We have reinterpreted the field redefinition geometrically as a change of coordinates in AdS$_5$ and shown that the conformal Galileons in the two representations are mapped into each other. 
We have also found the explicit form of the mapping and its inverse, given in (\ref{Dil2AdS}) and  (\ref{AdS2Dil}).
The knowledge of the map also allowed us to explicitly check the equivalence of the $2\rightarrow 2$ dilaton scattering at low energies in the two representations.
Notice that the equivalence requires that we keep the whole maps (\ref{Dil2AdS}) and  (\ref{AdS2Dil}) and we cannot truncate them: the leading terms in one representation are mapped into all the Galileons in the other representation and we need all of them to get the same S-matrix.

The mapping becomes more interesting and subtle when applied to non-trivial backgrounds. In particular, we have shown that in a class of 
backgrounds essentially invariant under the mapping (of which the Genesis scenario is a specific case  \cite{Nicolis:2009qm,Creminelli:2010ba,Hinterbichler:2012yn}), luminal fluctuations in the Weyl representation
are mapped to strictly sub-luminal fluctuations in the DBI representation, where luminality is measured with respect to the Minkowski light-cone.
The luminal Weyl fluctuations are instead mapped to DBI luminal fluctuations, if in the second case luminality is measured with respect to the induced brane metric.
This result seems to indicate that even when gravity is decoupled, the criterion of luminality around a Minkowski light-cone is in general not  well-defined.
The key question is now: given that the DBI and Weyl representations are different IR descriptions of the same physical system, how should we interpret these results? 
If superluminality (again, with respect to the Minkowski flat metric) appears in one description but not in the other, is the existence of a local and causal UV completion of this system ruled out?
We do not have a firm answer to this question, that deserves further work. 

\subsection*{Acknowledgements}
It is a pleasure to thank A.~Joyce, R. Rattazzi, M. Simonovi\'c, G.~Trevisan, G.~Villadoro, and especially  A.~Nicolis for useful comments. 
The work of ET is supported in part by the European Programme UNILHC contract PITN-GA-2009-237920 and by MIUR-FIRB grant RBFR12H1MW .


\footnotesize
\parskip 0pt

\begin{thebibliography}{99}

\bibitem{Bellucci:2002ji} 
  S.~Bellucci, E.~Ivanov and S.~Krivonos,
  ``AdS / CFT equivalence transformation,''
  Phys.\ Rev.\ D {\bf 66}, 086001 (2002)
  [Erratum-ibid.\ D {\bf 67}, 049901 (2003)]
  [hep-th/0206126].

\bibitem{Coleman:1969sm} 
  S.~R.~Coleman, J.~Wess and B.~Zumino,
  ``Structure of phenomenological Lagrangians. 1.,''
  Phys.\ Rev.\  {\bf 177}, 2239 (1969).

\bibitem{Callan:1969sn} 
  C.~G.~Callan, Jr., S.~R.~Coleman, J.~Wess and B.~Zumino,
  ``Structure of phenomenological Lagrangians. 2.,''
  Phys.\ Rev.\  {\bf 177}, 2247 (1969).
  
\bibitem{Volkov:1973vd} 
  D.~V.~Volkov,
  ``Phenomenological Lagrangians,''
  Fiz.\ Elem.\ Chast.\ Atom.\ Yadra {\bf 4}, 3 (1973).
  
\bibitem{Ivanov:1975zq} 
  E.~A.~Ivanov and V.~I.~Ogievetsky,
  ``The Inverse Higgs Phenomenon in Nonlinear Realizations,''
  Teor.\ Mat.\ Fiz.\  {\bf 25}, 164 (1975).
  
\bibitem{Nicolis:2008in} 
  A.~Nicolis, R.~Rattazzi and E.~Trincherini,
  ``The Galileon as a local modification of gravity,''
  Phys.\ Rev.\ D {\bf 79}, 064036 (2009)
  [arXiv:0811.2197 [hep-th]].
  
\bibitem{deRham:2010eu} 
  C.~de Rham and A.~J.~Tolley,
  ``DBI and the Galileon reunited,''
  JCAP {\bf 1005}, 015 (2010)
  [arXiv:1003.5917 [hep-th]].
  
\bibitem{Nicolis:2009qm} 
  A.~Nicolis, R.~Rattazzi and E.~Trincherini,
  ``Energy's and amplitudes' positivity,''
  JHEP {\bf 1005}, 095 (2010)
  [Erratum-ibid.\  {\bf 1111}, 128 (2011)]
  [arXiv:0912.4258 [hep-th]].
  
\bibitem{Creminelli:2010ba} 
  P.~Creminelli, A.~Nicolis and E.~Trincherini,
  ``Galilean Genesis: An Alternative to inflation,''
  JCAP {\bf 1011}, 021 (2010)
  [arXiv:1007.0027 [hep-th]].
    
\bibitem{Hinterbichler:2012yn} 
  K.~Hinterbichler, A.~Joyce, J.~Khoury and G.~E.~J.~Miller,
  ``DBI Genesis: An Improved Violation of the Null Energy Condition,''
  arXiv:1212.3607 [hep-th].
  
   \bibitem{Adams:2006sv}
  A.~Adams, N.~Arkani-Hamed, S.~Dubovsky, A.~Nicolis and R.~Rattazzi,
  ``Causality, analyticity and an IR obstruction to UV completion,''
  JHEP {\bf 0610} (2006) 014
  [hep-th/0602178].
    
   \bibitem{Maldacena:1997re}
  J.~M.~Maldacena,
  ``The Large N limit of superconformal field theories and supergravity,''
  Adv.\ Theor.\ Math.\ Phys.\  {\bf 2} (1998) 231
  [hep-th/9711200].
  
\bibitem{Rattazzi:2000hs} 
  R.~Rattazzi and A.~Zaffaroni,
  ``Comments on the holographic picture of the Randall-Sundrum model,''
  JHEP {\bf 0104}, 021 (2001)
  [hep-th/0012248].
  
\bibitem{Elvang:2012st} 
  H.~Elvang, D.~Z.~Freedman, L.~-Y.~Hung, M.~Kiermaier, R.~C.~Myers and S.~Theisen,
  ``On renormalization group flows and the a-theorem in 6d,''
  JHEP {\bf 1210}, 011 (2012)
  [arXiv:1205.3994 [hep-th]].
 
\bibitem{Creminelli:2012my} 
  P.~Creminelli, K.~Hinterbichler, J.~Khoury, A.~Nicolis and E.~Trincherini,
  ``Subluminal Galilean Genesis,''
  JHEP {\bf 1302}, 006 (2013)
  [arXiv:1209.3768 [hep-th]].
  
\bibitem{Goon:2011qf} 
  G.~Goon, K.~Hinterbichler and M.~Trodden,
  ``Symmetries for Galileons and DBI scalars on curved space,''
  JCAP {\bf 1107}, 017 (2011)
  [arXiv:1103.5745 [hep-th]].

   
\bibitem{Komargodski:2011vj} 
  Z.~Komargodski and A.~Schwimmer,
  ``On Renormalization Group Flows in Four Dimensions,''
  JHEP {\bf 1112}, 099 (2011)
  [arXiv:1107.3987 [hep-th]].
    
   
\bibitem{Goon:2012dy}
  G.~Goon, K.~Hinterbichler, A.~Joyce and M.~Trodden,
  ``Galileons as Wess-Zumino Terms,''
  JHEP {\bf 1206} (2012) 004
  [arXiv:1203.3191 [hep-th]].
  
\bibitem{Hinterbichler:2012fr} 
  K.~Hinterbichler, A.~Joyce, J.~Khoury and G.~E.~J.~Miller,
  ``DBI Realizations of the Pseudo-Conformal Universe and Galilean Genesis Scenarios,''
  JCAP {\bf 1212}, 030 (2012)
  [arXiv:1209.5742 [hep-th]].
  
\bibitem{Deffayet:2009wt} 
  C.~Deffayet, G.~Esposito-Farese and A.~Vikman,
  ``Covariant Galileon,''
  Phys.\ Rev.\ D {\bf 79}, 084003 (2009)
  [arXiv:0901.1314 [hep-th]].
  
  
\end{thebibliography}
\end{document}